\documentstyle[12pt,epsf]{article}
\newcommand{\logqmms}{l_{qm}}

\newcommand{\logmsos}{L_{ms}}

\newcommand{\logqmums}{l_{q\mu}}

\newcommand{\gsim}{\;\rlap{\lower 3.5 pt \hbox{$\mathchar \sim$}} \raise 1pt
 \hbox {$>$}\;}
\newcommand{\lsim}{\;\rlap{\lower 3.5 pt \hbox{$\mathchar \sim$}} \raise 1pt
 \hbox {$<$}\;}

\renewcommand{\thefootnote}{\fnsymbol{footnote}}
\begin{document}    

\title{\vskip-3cm{\baselineskip14pt
\centerline{\normalsize\hfill MPI/PhT/97--66}
\centerline{\normalsize\hfill TTP97--40\footnote{The 
  complete postscript file of this
  preprint, including figures, is available via anonymous ftp at
  www-ttp.physik.uni-karlsruhe.de (129.13.102.139) as /ttp97-40/ttp97-40.ps 
  or via www at http://www-ttp.physik.uni-karlsruhe.de/cgi-bin/preprints.}}
\centerline{\normalsize\hfill hep-ph/9710413}
\centerline{\normalsize\hfill October 1997}
}
\vskip1.5cm
${\cal O}(\alpha_s^2)$ Corrections to Top Quark Production\\
at $e^+e^-$ Colliders
}
\author{
 R.~Harlander$^{a}$\thanks{Supported by the ``Landesgraduiertenf\"orderung'' 
                           at the University of Karlsruhe.}
 and 
 M.~Steinhauser$^{b}$
}
\date{}
\maketitle

\begin{center}
$^a${\it Institut f\"ur Theoretische Teilchenphysik,
    Universit\"at Karlsruhe,\\ D-76128 Karlsruhe, Germany,\\ }
  \vspace{3mm}
$^b${\it Max-Planck-Institut f\"ur Physik,
    Werner-Heisenberg-Institut,\\ D-80805 Munich, Germany.\\ }
\end{center}

\begin{abstract}
  \noindent 
In this article we evaluate mass corrections up to ${\cal O}((m^2/q^2)^6)$
to the three-loop polarization function induced by an axial-vector
current. Special emphasis is put on the evaluation of the singlet diagram
which is absent in the vector case.
As a physical application ${\cal O}(\alpha_s^2)$ corrections to the 
production of top quarks at future $e^+e^-$ colliders is considered.
It is demonstrated that for center of mass energies 
$\sqrt{s}\gsim500$~GeV the inclusion of the first seven terms 
into the cross section leads to 
a reliable description.

%
% 12.38.-t Quantum chromodynamics
% 12.38.Bx Perturbative calculations
% 13.85.Lg Total cross sections
% 14.65.Ha Top quarks
%
\end{abstract}

%\thispagestyle{empty}
%\newpage
%\setcounter{page}{1}

%%%%%%%%%%%%%%%%%%%%%%%%%%%%%%%%%%%%%%%%%%%%%%%%%%%%%%%%%%%%
%%%%%%%%%%%%%%%%%%%%%%%%%%%%%%%%%%%%%%%%%%%%%%%%%%%%%%%%%%%%

\renewcommand{\thefootnote}{\arabic{footnote}}
\setcounter{footnote}{0}

In the total cross section $\sigma(e^+e^-\to\mbox{hadrons})$
corrections arising from the finite mass, $m$, of the produced 
quarks may often be neglected.
Concerning precision measurements around the $Z$ resonance
first order mass corrections, known up to ${\cal O}(\alpha_s^3)$
\cite{CKKRep,CheKue96},
are usually adequate.
However, having in mind top quark production at future colliders like 
the NLC with a center of mass energy of $\sqrt{s}=500$~GeV
higher order terms in $m^2/s$ may become important.
The velocity of the produced particles 
is then $v\approx 0.7$ which means that on one side
threshold effects are not important and on the other side
we are not in the region of very high energies.

In Ref.~\cite{CheHoaKueSteTeu96} the contribution of the 
photon to the production of top quarks was considered.
In this article also the exchange of the $Z$ boson is included.
Hence, in a first step results for the axial-vector polarization 
function up to ${\cal O}(\alpha_s^2)$ are presented. 
The imaginary part in combination with the
recently evaluated rate for the vector case 
\cite{CheHarKueSte97}
directly leads to the cross section
$\sigma(e^+e^-\to t\bar{t} + X)$ 
mediated by a virtual $Z$ boson.
The ${\cal O}(\alpha_s)$ corrections to this process were considered in
\cite{JerLaeZer82}.

\begin{figure}[t]
 \begin{center}
 \begin{tabular}{cc}
   \leavevmode
   \epsfxsize=6.5cm
   \epsffile[131 314 481 478]{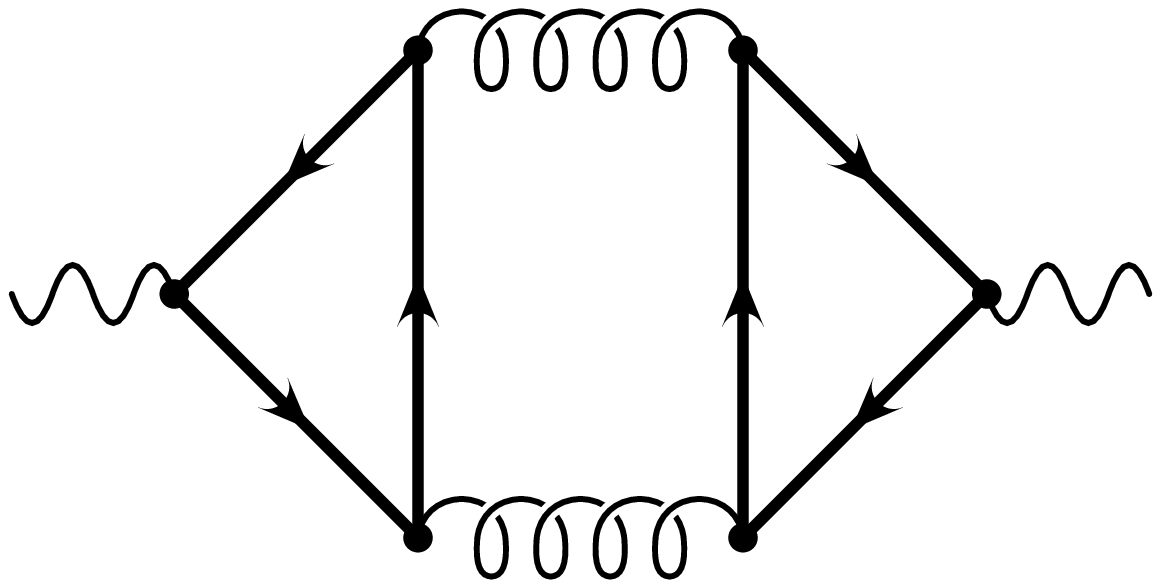}
   &
   \epsfxsize=6.5cm
   \epsffile[131 314 481 478]{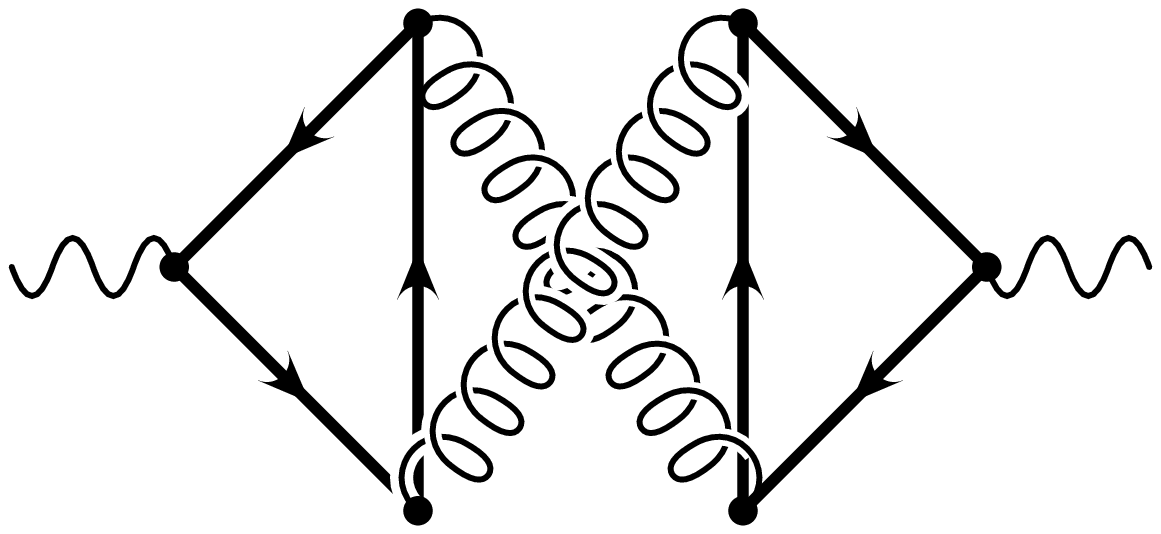} 
 \end{tabular}
 \caption{\label{figdiasing}Diagrams contributing to $\Pi_S^{a}$.
          In the triangle loops either a top or bottom quark may be present.
          }
 \end{center}
\end{figure}

To be more precise let us define the axial-vector current correlator as:
\begin{eqnarray}
\left(-q^2g_{\mu\nu}+q_\mu q_\nu\right)\,\Pi^a(q^2)
+q_\mu q_\nu\,\Pi^a_L(q^2)
&=&
i\int dx\,e^{iqx}\langle 0|Tj^a_\mu(x) j^a_\nu(0)|0 \rangle
\label{eqpivadef}
\end{eqnarray}
with 
$
j_\mu^a = \bar{\psi}\gamma_\mu\gamma_5 \psi.
$
In the following we will only present results for $\Pi^a(q^2)$\footnote{
The longitudinal part, $\Pi_L^a(q^2)$, of the non-singlet contribution, e.g.,
is via the axial Ward identity
directly connected to the pseudo-scalar polarization function $\Pi^p(q^2)$,
for which the high energy expansion was considered in
\cite{HarSte97}.}.
It is convenient to write
\begin{eqnarray}
\Pi^a(q^2) &=& \Pi^{(0),a}(q^2) 
         + \frac{\alpha_s(\mu^2)}{\pi} C_F \Pi^{(1),a}(q^2)
         + \left(\frac{\alpha_s(\mu^2)}{\pi}\right)^2\Pi^{(2),a}(q^2)
         + \ldots\,\,,
\nonumber\\
\Pi^{(2),a} &=&
                C_F^2       \Pi_A^{(2),a}
              + C_A C_F     \Pi_{\it NA}^{(2),a}
              + C_F T   n_l \Pi_l^{(2),a}
              + C_F T       \Pi_F^{(2),a},
\label{eqpi2}
\end{eqnarray}
with the $SU(3)$ colour factors $C_F=4/3, C_A=3$ and $T=1/2$. 
$\Pi_A^{(2),a}$ is the abelian contribution already present in QED and
$\Pi_{NA}^{(2),a}$ originates from the non-abelian structure
specific for QCD. The polarization functions containing a second
massless or massive quark loop are denoted 
by $\Pi_l^{(2),a}$ and $\Pi_F^{(2),a}$, respectively.
$\Pi^a$ represents the so-called non-singlet part. However, for
external axial-vector currents already at 
${\cal O}(\alpha_s^2)$ there exists also a singlet or
double-triangle contribution:
\begin{eqnarray}
\Pi_S^a(q^2) &=& 
\left(\frac{\alpha_s(\mu^2)}{\pi}\right)^2\,C_FT\,\Pi_S^{(2),a}(q^2).
\label{eqpiS2}
\end{eqnarray} 
As $\Pi_S^a$ depends on the properties of both members of the 
fermion doublet we will from now on specify to the top-bottom case.
The generalization to other quark flavours is obvious. 
For this contribution it is convenient to replace the current $j^a_\mu$
in Eq.~(\ref{eqpivadef}) by 
$
\bar{t}\gamma_\mu\gamma_5 t - \bar{b}\gamma_\mu\gamma_5 b
$
because in this combination the axial anomaly cancels.
In Fig.~\ref{figdiasing} the relevant diagrams are depicted.

Similar relations as in Eqs.~(\ref{eqpi2}) and (\ref{eqpiS2}) 
also hold for $R^a(s)$ and $R_S^a(s)$, respectively, defined through
\begin{eqnarray}
R_{(S)}^a (s)&=&12\pi\,\mbox{Im}\,\Pi_{(S)}^a(q^2=s+i\epsilon),
\end{eqnarray}
so that the cross section for the inclusive production of top quarks
may be written as
\begin{eqnarray}
%%%\lefteqn{
R_t(s)
&=&
%%%\,\,=\,\,
\frac{\sigma(e^+e^-\to t\bar{t} + X)}{\sigma_{pt}}
\nonumber\\
&=&
(v_e^2+a_e^2)a_t^2 \left(\frac{s}{s-M_Z^2}\right)^2 
\,\,\left(R^a(s) + R_S^a(s)
- \left(\frac{\alpha_s}{\pi}\right)^2C_F T R^{(2),a}_{Sb}(s)\right)
%%%}
\nonumber\\&&\mbox{}
+
\bigg[
  Q_e^2Q_t^2 
+ 2 Q_e v_e Q_t v_t \frac{s}{s-M_Z^2}
%%%\nonumber\\&&\mbox{}
+ (v_e^2+a_e^2)v_t^2 \left(\frac{s}{s-M_Z^2}\right)^2 
\bigg]\,\,R^v(s),
\label{eqtt}
\end{eqnarray}
with
$\sigma_{pt}=4\pi\alpha^2/3s$,
$v_f=(I_3^f-2Q_fs_\theta^2)/(2s_\theta c_\theta)$,
$a_f=I_3^f/(2s_\theta c_\theta)$,
$Q_e=-1, Q_t=2/3, I_3^e=-1/2$ and $I_3^t=1/2$.
Furthermore we have $c_\theta^2=1-s_\theta^2$ with $s_\theta$
being the sine of the weak mixing angle.
$R^v(s)$ is given in 
\cite{CheHarKueSte97}
and both $R^a(s)$ and $R_S^a(s)$ will be presented below.
$R_{Sb}^{(2),a}$ is the contribution from cuts of the singlet diagram
that do not involve top quarks.
Non-singlet contributions with the photon or 
$Z$ boson coupling to a light quark flavour
and the top quarks produced via gluon splitting 
\cite{HoaJezKueTeu94} will be neglected 
as their numerical values are tiny 
\cite{CheHoaKueSteTeu96}.

The computation of $\Pi^a$ naturally splits into two parts:
Firstly into the non-singlet contribution 
where the anticommuting  
definition of $\gamma_5$ may be used. Here the calculation of the
diagrams is in close analogy to the vector case. Hence
we refer for details to 
\cite{CheHarKueSte97}.

The second part, the singlet contribution $\Pi_S^a$,
is connected with the axial anomaly and is
not present in $\Pi^v$. 
Let us briefly describe our treatment of these diagrams.
Actually three graphs have to be considered, namely the cases when
two top quarks, one top and one bottom quark or two bottom quarks 
are running in the triangle loops. 
One may argue that the last combination only contributes to
the cross section into bottom quarks which is not the process
under consideration. However, only the proper combination of all
three parts guarantees the cancellation of the anomaly.
From the final result the cuts arising from bottom quarks
have to be subtracted, of course.

For the evaluation of $\Pi_S^{(2),a}$ naive $\gamma_5$ fails to work.
We follow the treatment introduced in
\cite{tHoVel72}
and formalized in
\cite{BreMai77}
and replace both axial-vector vertices 
according to
\cite{AkyDel73}
\begin{eqnarray}
\gamma_\mu\gamma_5 &\to& 
\frac{i}{3!}\,\epsilon_{\mu\lambda\rho\sigma}\gamma^{[\lambda\rho\sigma]},
\end{eqnarray}
where $\gamma^{[\lambda\rho\sigma]}$ is the antisymmetric combination
of three $\gamma$ matrices which can be written as
$
\gamma^{[\lambda\rho\sigma]}=(
\gamma^\lambda \gamma^\rho \gamma^\sigma
-
\gamma^\sigma \gamma^\rho \gamma^\lambda
)/2$.
In a first step the $\epsilon$-tensors are put aside and 
the new object with six external indices, 
$\Pi^{[\lambda\rho\sigma]}_{[\lambda^\prime\rho^\prime\sigma^\prime]}$,
defined through
\begin{eqnarray}
\Pi_{\mu\nu} = \left(\frac{i}{3!}\right)^2 
\epsilon_{\mu\lambda\rho\sigma} \,
\epsilon_{\nu}{}^{\lambda^\prime\rho^\prime\sigma^\prime} \,
\Pi^{[\nu\rho\sigma]}_{[\nu^\prime\rho^\prime\sigma^\prime]},
\end{eqnarray}
is treated until
the momentum integration and renormalization is done
and a finite quantity is available \cite{Lar93}. Then the contraction 
with the $\epsilon$-tensors is performed.
It is possible to show that the contribution from the singlet diagrams
may be computed from the relation
\cite{CheKwi932}
\begin{eqnarray}
\Pi^a_S(q^2) &=& -\frac{
q_{\sigma}q^{\sigma^\prime}
\Pi^{[\lambda\rho\sigma]}_{[\lambda\rho\sigma^\prime]}
}{6(q^2)^2},
\end{eqnarray}
which means that we can treat the scalar quantity 
$q_{\sigma}q^{\sigma^\prime}\Pi^{[\nu\rho\sigma]}_{[\nu\rho\sigma^\prime]}$
in complete analogy to the non-singlet diagrams.
We should mention that a finite renormalization of the singlet 
axial-vector current 
\cite{Tru79}
has not to be performed in the order considered in this paper.

Using the large momentum procedure the first seven terms
in the $m^2/q^2$-expansion of $\Pi^a(q^2)$ have been evaluated. 
We refrain from listing the results separated into the 
contributions from the different colour factors and
present the results for the proper sum keeping only $n_l$, the
number of light (massless) quarks, as arbitrary parameter
($\logqmms \equiv \ln(-q^2/m_t^2)$, $\logqmums \equiv \ln(-q^2/\mu^2)$):
\begin{eqnarray}
  \bar{\Pi}^{(0),a} &=& {3\over 16\pi^2} \bigg\{
         {20\over 9} 
       - {4\over 3}\,\logqmums 
    + {m_t^2\over q^2}\,\left(
           - 8 
           + 8\,\logqmums
          \right) 
    + \left({m_t^2\over q^2}\right)^{2}\,\left(
           - 12 
           - 8\,\logqmms
          \right) 
       \nonumber\\&&\mbox{}
    + \left({m_t^2\over q^2}\right)^{3}\,\left(
             {8\over 9} 
           - {16\over 3}\,\logqmms
          \right) 
    + \left({m_t^2\over q^2}\right)^{4}\,\left(
             {14\over 3} 
           - 8\,\logqmms
          \right)
       \nonumber\\&&\mbox{}
    + \left({m_t^2\over q^2}\right)^{5}\,\left(
             {188\over 15} 
           - 16\,\logqmms
          \right) 
    + \left({m_t^2\over q^2}\right)^{6}\,\left(
             {1516\over 45} 
           - {112\over 3}\,\logqmms
          \right)
\bigg\}
+ \ldots\,\,,
\label{eqpi0a}
\\
%%%%%%%%%%%%%%%%%%%%%%%%%%%%%%%%%%%%%%%
  \bar{\Pi}^{(1),a} &=& {3\over 16\pi^2} \bigg\{
             {55\over 12} 
           - 4\,\zeta_3 
           - \logqmums 
           \nonumber\\&&\mbox{}
    + {m_t^2\over q^2}\,\bigg[
               - {107\over 2} 
               + 24\,\zeta_3
               + 22\,\logqmums 
               - 6\,\logqmums^2 
              \bigg]
           \nonumber\\&&\mbox{}
    + \left({m_t^2\over q^2}\right)^{2}\,\bigg[
                 {2\over 3} 
               - 32\,\zeta_3
               - 34\,\logqmms 
               - 12\,\logqmms^2 
               + \left(
                     24 
                   + 24\,\logqmms
                  \right)\,\logqmums 
              \bigg]
           \nonumber\\&&\mbox{}
    + \left({m_t^2\over q^2}\right)^{3}\,\bigg[
               - {304\over 27} 
               - {868\over 27}\,\logqmms 
               - {160\over 9}\,\logqmms^2 
               + \left(
                   - 12 
                   + 24\,\logqmms
                  \right)\,\logqmums
              \bigg] 
           \nonumber\\&&\mbox{}
    + \left({m_t^2\over q^2}\right)^{4}\,\bigg[
                 {5671\over 216} 
               - {449\over 9}\,\logqmms 
               - 33\,\logqmms^2 
               + \left(
                   - 40 
                   + 48\,\logqmms
                  \right)\,\logqmums
              \bigg]
           \nonumber\\&&\mbox{}
    + \left({m_t^2\over q^2}\right)^{5}\,\bigg[
                 {1718971\over 13500} 
               - {77954\over 675}\,\logqmms 
               - {3922\over 45}\,\logqmms^2 
               + \left(
                   - 118 
                   + 120\,\logqmms
                  \right)\,\logqmums
              \bigg]
           \nonumber\\&&\mbox{}
    + \left({m_t^2\over q^2}\right)^{6}\,\bigg[
                 {9302591\over 20250} 
               - {193546\over 675}\,\logqmms 
               - {11308\over 45}\,\logqmms^2 
       \nonumber\\&&\mbox{\hspace{.5cm}}
               + \left(
                   - {1796\over 5} 
                   + 336\,\logqmms
                  \right)\,\logqmums
              \bigg]
\bigg\}
+ \ldots\,\,,
\label{eqpi1a}
\\
%%%%%%%%%%%%%%%%%%%%%%%%%%%%%%%%%%%%%%%
  \bar{\Pi}^{(2),a} &=& {3\over 16\pi^2} \bigg\{
         {118379\over 1944} 
       - {1582\over 27}\,\zeta_3 
       + {100\over 9}\,\zeta_5
       + \left(
           - {343\over 18} 
           + {124\over 9}\,\zeta_3
          \right)\,\logqmums 
       + {31\over 18}\,\logqmums^2 
       \nonumber\\&&\mbox{\hspace{.5cm}}
       + n_l \left(
           - {3701\over 972} 
           + {76\over 27}\,\zeta_3
           + \left(
                 {11\over 9} 
               - {8\over 9}\,\zeta_3
              \right)\,\logqmums 
           - {1\over 9}\,\logqmums^2 
          \right)
       \nonumber\\&&\mbox{}
    + {m_t^2\over q^2}\,\bigg[
           - {18973\over 27} 
           + {4612\over 9}\,\zeta_3 
           + 2\,\zeta_4 
           - 220\,\zeta_5
       \nonumber\\&&\mbox{\hspace{.5cm}}
           + \left(
                 {7919\over 18} 
               - {452\over 3}\,\zeta_3
              \right)\,\logqmums 
           - {898\over 9}\,\logqmums^2 
           + {110\over 9}\,\logqmums^3 
       \nonumber\\&&\mbox{\hspace{.5cm}}
           + n_l \left(
                 {857\over 27} 
               - {128\over 9}\,\zeta_3 
               + \left(
                   - {151\over 9} 
                   + {16\over 3}\,\zeta_3
                  \right)\,\logqmums
               + {32\over 9}\,\logqmums^2 
               - {4\over 9}\,\logqmums^3 
              \right) 
          \bigg] 
       \nonumber\\&&\mbox{}
    + \left({m_t^2\over q^2}\right)^{2}\,\bigg[
             {8615\over 162} 
           - {9140\over 27}\,\zeta_3 
           - {32\over 3}\,\zeta_4 
           - {3080\over 27}\,\zeta_5
           + {16\over 9}\,B_4 
       \nonumber\\&&\mbox{\hspace{.5cm}}
           + \left(
               - {10987\over 27} 
               + {32\over 3}\,\zeta_3
              \right)\,\logqmms 
           - {1430\over 9}\,\logqmms^2 
           - {316\over 9}\,\logqmms^3 
       \nonumber\\&&\mbox{\hspace{.5cm}}
           + \left(
                 {910\over 27} 
               + {2528\over 9}\,\zeta_3
               + {1094\over 3}\,\logqmms 
               + {316\over 3}\,\logqmms^2 
              \right)\,\logqmums 
           + \left(
               - {220\over 3} 
               - {316\over 3}\,\logqmms
              \right)\,\logqmums^2 
       \nonumber\\&&\mbox{\hspace{.5cm}}
           + n_l \Bigg(
               - {149\over 81} 
               + {416\over 27}\,\zeta_3
               + {362\over 27}\,\logqmms 
               + {40\over 9}\,\logqmms^2 
               + {8\over 9}\,\logqmms^3 
       \nonumber\\&&\mbox{\hspace{.5cm}}
               + \left(
                   - {116\over 27} 
                   - {64\over 9}\,\zeta_3
                   - 12\,\logqmms 
                   - {8\over 3}\,\logqmms^2 
                  \right)\,\logqmums 
               + \left(
                     {8\over 3} 
                   + {8\over 3}\,\logqmms
                  \right)\,\logqmums^2 
              \Bigg)
          \bigg] 
       \nonumber\\&&\mbox{}
    + \left({m_t^2\over q^2}\right)^{3}\,\bigg[
             {748169\over 26244} 
           - {18718\over 81}\,\zeta_3 
           - {64\over 9}\,\zeta_4 
           - {920\over 27}\,\zeta_5
           + {32\over 27}\,B_4 
       \nonumber\\&&\mbox{\hspace{.5cm}}
           + \left(
               - {639715\over 1458} 
               - {976\over 27}\,\zeta_3
              \right)\,\logqmms 
           - {66698\over 243}\,\logqmms^2 
           - {55064\over 729}\,\logqmms^3 
       \nonumber\\&&\mbox{\hspace{.5cm}}
           + \left(
               - {5342\over 243} 
               + {98008\over 243}\,\logqmms 
               + {16480\over 81}\,\logqmms^2
              \right)\,\logqmums 
           + \left(
                 {302\over 3} 
               - {412\over 3}\,\logqmms
              \right)\,\logqmums^2 
       \nonumber\\&&\mbox{\hspace{.5cm}}
           + n_l \Bigg(
               - {3167\over 2187} 
               + {224\over 27}\,\zeta_3
               + {10630\over 729}\,\logqmms 
               + {512\over 81}\,\logqmms^2 
               + {176\over 243}\,\logqmms^3 
       \nonumber\\&&\mbox{\hspace{.5cm}}
               + \left(
                   - {68\over 243} 
                   - {2816\over 243}\,\logqmms 
                   - {320\over 81}\,\logqmms^2
                  \right)\,\logqmums 
               + \left(
                   - {4\over 3} 
                   + {8\over 3}\,\logqmms
                  \right)\,\logqmums^2 
              \Bigg) 
          \bigg] 
       \nonumber\\&&\mbox{}
    + \left({m_t^2\over q^2}\right)^{4}\,\bigg[
             {88895269\over 209952} 
           - {66964\over 243}\,\zeta_3 
           - {32\over 3}\,\zeta_4 
           - {560\over 27}\,\zeta_5
           + {16\over 9}\,B_4 
       \nonumber\\&&\mbox{\hspace{.5cm}}
           + \left(
               - {14315023\over 17496} 
               + {64\over 9}\,\zeta_3
              \right)\,\logqmms 
           - {422909\over 729}\,\logqmms^2 
           - {122420\over 729}\,\logqmms^3 
       \nonumber\\&&\mbox{\hspace{.5cm}}
           + \left(
               - {1400761\over 1944} 
               + {63863\over 81}\,\logqmms 
               + {1397\over 3}\,\logqmms^2
              \right)\,\logqmums 
           + \left(
                 {3116\over 9} 
               - {1016\over 3}\,\logqmms
              \right)\,\logqmums^2 
       \nonumber\\&&\mbox{\hspace{.5cm}}
           + n_l \Bigg(
               - {65785\over 5832} 
               + {80\over 9}\,\zeta_3
               + {12431\over 486}\,\logqmms 
               + {671\over 54}\,\logqmms^2 
               + {38\over 27}\,\logqmms^3 
       \nonumber\\&&\mbox{\hspace{.5cm}}
               + \left(
                     {12871\over 972} 
                   - {1618\over 81}\,\logqmms 
                   - {22\over 3}\,\logqmms^2
                  \right)\,\logqmums 
               + \left(
                   - {40\over 9} 
                   + {16\over 3}\,\logqmms
                  \right)\,\logqmums^2 
              \bigg) 
          \bigg] 
       \nonumber\\&&\mbox{}
    + \left({m_t^2\over q^2}\right)^{5}\,\bigg[
             {1098529906403\over 524880000} 
           - {4485269\over 12150}\,\zeta_3 
           - {64\over 3}\,\zeta_4 
           - {1120\over 27}\,\zeta_5
           + {32\over 9}\,B_4 
       \nonumber\\&&\mbox{\hspace{.5cm}}
           + \left(
               - {1102325809\over 540000} 
               + {148\over 3}\,\zeta_3
              \right)\,\logqmms 
           - {1318561453\over 729000}\,\logqmms^2 
           - {9340049\over 18225}\,\logqmms^3 
       \nonumber\\&&\mbox{\hspace{.5cm}}
           + \left(
               - {374774041\over 121500} 
               + {12902714\over 6075}\,\logqmms 
               + {592222\over 405}\,\logqmms^2
              \right)\,\logqmums 
       \nonumber\\&&\mbox{\hspace{.5cm}}
           + \left(
                 {10349\over 9} 
               - {3020\over 3}\,\logqmms
              \right)\,\logqmums^2 
       \nonumber\\&&\mbox{\hspace{.5cm}}
           + n_l \Bigg(
               - {124701659\over 2733750} 
               + {1376\over 135}\,\zeta_3
               + {1173494\over 18225}\,\logqmms 
               + {68779\over 2025}\,\logqmms^2 
               + {4244\over 1215}\,\logqmms^3 
       \nonumber\\&&\mbox{\hspace{.5cm}}
               + \left(
                     {3046471\over 60750} 
                   - {290908\over 6075}\,\logqmms 
                   - {7844\over 405}\,\logqmms^2
                  \right)\,\logqmums 
               + \left(
                   - {118\over 9} 
                   + {40\over 3}\,\logqmms
                  \right)\,\logqmums^2 
              \Bigg) 
          \bigg] 
       \nonumber\\&&\mbox{}
    + \left({m_t^2\over q^2}\right)^{6}\,\bigg[
             {2399908800637\over 262440000} 
           - {366236\over 1215}\,\zeta_3 
           - {448\over 9}\,\zeta_4 
           - {1120\over 9}\,\zeta_5
           + {224\over 27}\,B_4 
       \nonumber\\&&\mbox{\hspace{.5cm}}
           + \left(
               - {114901711063\over 21870000} 
               + {7904\over 45}\,\zeta_3
              \right)\,\logqmms 
           - {4577019727\over 729000}\,\logqmms^2 
           - {659557\over 405}\,\logqmms^3 
       \nonumber\\&&\mbox{\hspace{.5cm}}
           + \left(
               - {424502089\over 36450} 
               + {7360838\over 1215}\,\logqmms 
               + {395780\over 81}\,\logqmms^2
              \right)\,\logqmums 
       \nonumber\\&&\mbox{\hspace{.5cm}}
           + \left(
                 {35462\over 9} 
               - {9800\over 3}\,\logqmms
              \right)\,\logqmums^2 
       \nonumber\\&&\mbox{\hspace{.5cm}}
           + n_l \Bigg(
               - {242108108\over 1366875} 
               + {5312\over 405}\,\zeta_3
               + {114707\over 675}\,\logqmms 
               + {623728\over 6075}\,\logqmms^2 
               + {11896\over 1215}\,\logqmms^3 
       \nonumber\\&&\mbox{\hspace{.5cm}}
               + \left(
                     {15364091\over 91125} 
                   - {765092\over 6075}\,\logqmms 
                   - {22616\over 405}\,\logqmms^2
                  \right)\,\logqmums 
       \nonumber\\&&\mbox{\hspace{.5cm}}
               + \left(
                   - {1796\over 45} 
                   + {112\over 3}\,\logqmms
                  \right)\,\logqmums^2 
              \Bigg) 
          \bigg]
\bigg\}
+ \ldots\,\,,
\label{eqpi2a}
\\
%%%%%%%%%%%%%%%%%%%%%%%%%%%%%%%%%%%%%%%
  \bar{\Pi}^{(2),a}_S &=& {3\over 16\pi^2} \bigg\{
    \left({m_t^2\over q^2}\right)^{2}\,\bigg[
           - {80\over 3}\,\zeta_3 
           + {320\over 3}\,\zeta_5
          \bigg]
       \nonumber\\&&\mbox{}
    + \left({m_t^2\over q^2}\right)^{3}\,\bigg[
             {380\over 3} 
           - 64\,\zeta_3
           + \left(
                 {296\over 3} 
               - 32\,\zeta_3
              \right)\,\logqmms 
           + 24\,\logqmms^2 
          \bigg] 
       \nonumber\\&&\mbox{}
    + \left({m_t^2\over q^2}\right)^{4}\,\bigg[
           - {3271\over 243} 
           - {416\over 9}\,\zeta_3 
           + \left(
                 {280\over 27} 
               + 32\,\zeta_3
              \right)\,\logqmms
           + {410\over 27}\,\logqmms^2 
           - {176\over 27}\,\logqmms^3 
          \bigg] 
       \nonumber\\&&\mbox{}
    + \left({m_t^2\over q^2}\right)^{5}\,\bigg[
           - {395921\over 2916} 
           - {5584\over 27}\,\zeta_3 
       \nonumber\\&&\mbox{\hspace{.5cm}}
           + \left(
                 {4111\over 54} 
               + {160\over 3}\,\zeta_3
              \right)\,\logqmms
           + {1340\over 9}\,\logqmms^2 
           - {1660\over 81}\,\logqmms^3 
          \bigg] 
       \nonumber\\&&\mbox{}
    + \left({m_t^2\over q^2}\right)^{6}\,\bigg[
           - {105441373\over 101250} 
           - {2420\over 3}\,\zeta_3 
       \nonumber\\&&\mbox{\hspace{.5cm}}
           + \left(
               - {6044237\over 40500} 
               + 112\,\zeta_3
              \right)\,\logqmms
           + {1177331\over 1350}\,\logqmms^2 
           - {15542\over 135}\,\logqmms^3 
          \bigg]
\bigg\}
+ \ldots\,\,,
\label{eqpi2as}
\end{eqnarray}
where $m_t$ is the $\overline{\mbox{MS}}$ top mass
and $\zeta$ is Riemann's zeta-function with the values
$\zeta_2=\pi^2/6$, $\zeta_3\approx1.20206$, $\zeta_4=\pi^4/90$ and
$\zeta_5\approx1.03693$.  $B_4\approx-1.76280$ 
is a constant typical for massive three-loop integrals
\cite{Bro92}.
The expansion of the two-loop quantity, $\bar{\Pi}^{(1),a}$
can be compared with the exact result
\cite{Kni90}. At order $\alpha_s^2$ the constant and quadratic 
terms are in agreement with 
\cite{CheKwi93,CheKueSte97}. 
Note that in the non-singlet contribution $m_t$ could be replaced 
by any other quark mass.
The singlet part, however,
gets modified if both quarks have to be considered as massive and 
even vanishes for a degenerate quark doublet. This is also the reason
for the absence of the first two terms in the
expansion for $m_t^2/q^2\to0$: For $m_t=0$ the top and bottom
quark are trivially degenerate and the contribution to the first 
order power corrections arise from a simple expansion of the diagrams
for small masses. According to the structure of the $\gamma$ matrices
from each triangle at least a factor $m_t^2$ has to come. This means that 
the $m_t^2/q^2$ corrections from the diagrams with two top triangles
cancel against the one with a top and a bottom triangle which has
an overall factor of two.

Taking the imaginary part of Eqs.~(\ref{eqpi0a}-\ref{eqpi2as})
and transforming the result into the
on-shell scheme concerning the top mass
\cite{GraBroGraSch90}
leads to
($\logmsos \equiv \ln(M_t^2/s)$):
\begin{eqnarray}
R^{(0),a} &=&
3\bigg\{
 1 - 6\frac{M_t^2}{s} + 6\left(\frac{M_t^2}{s}\right)^2
+ 4\left(\frac{M_t^2}{s}\right)^3
       \nonumber\\&&\mbox{}
+ 6\left(\frac{M_t^2}{s}\right)^4
+ 12\left(\frac{M_t^2}{s}\right)^5
+ 28\left(\frac{M_t^2}{s}\right)^6
\bigg\}
+ \ldots\,\,,
\\
%%%%%%%%%%%%%%%%%%%%%%%%%%%%%%%%%%%%%%
  R^{(1),a} &=& 3\bigg\{
         {3\over 4} 
    + {M_t^2\over s}\,\left(
           - {9\over 2} 
           - 9\,\logmsos
          \right)
    + \left({M_t^2\over s}\right)^{2}\,\left(
           - {33\over 2} 
           + 18\,\logmsos
          \right)
       \nonumber\\&&\mbox{}
    + \left({M_t^2\over s}\right)^{3}\,\left(
             {82\over 9} 
           + {28\over 3}\,\logmsos
          \right)
    + \left({M_t^2\over s}\right)^{4}\,\left(
             {233\over 12} 
           + {45\over 2}\,\logmsos
          \right)
       \nonumber\\&&\mbox{}
    + \left({M_t^2\over s}\right)^{5}\,\left(
             {12401\over 225} 
           + {739\over 15}\,\logmsos
          \right)
    + \left({M_t^2\over s}\right)^{6}\,\left(
             {66803\over 450} 
           + {1906\over 15}\,\logmsos
          \right)
\bigg\}
+ \ldots\,\,,
\\
%%%%%%%%%%%%%%%%%%%%%%%%%%%%%%%%%
  R^{(2),a} &=& 3\bigg\{
         {343\over 24} 
       - {31\over 3}\,\zeta_3 
       + n_l\,\left(
           - {11\over 12} 
           + {2\over 3}\,\zeta_3
          \right) 
       \nonumber\\&&\mbox{}
    + {M_t^2\over s}\,\bigg[
           - {937\over 6} 
           + \left(
                 79 
               + 8\,\ln 2
              \right)\,\zeta_2 
           + 111\,\zeta_3
           - {613\over 6}\,\logmsos 
           + {7\over 2}\,\logmsos^2 
       \nonumber\\&&\mbox{\hspace{.5cm}}
           + n_l\,\left(
                 {20\over 3} 
               - 6\,\zeta_2 
               - 4\,\zeta_3
               + {13\over 3}\,\logmsos 
               - \logmsos^2 
              \right) 
          \bigg]
       \nonumber\\&&\mbox{}
    + \left({M_t^2\over s}\right)^{2}\,\bigg[
             39 
           + \left(
               - 206 
               - 16\,\ln 2
              \right)\,\zeta_2 
           - {644\over 3}\,\zeta_3 
           + {255\over 2}\,\logmsos 
           + 17\,\logmsos^2 
       \nonumber\\&&\mbox{\hspace{.5cm}}
           + n_l\,\left(
                 5 
               + 12\,\zeta_2 
               + {16\over 3}\,\zeta_3
               - 11\,\logmsos 
               + 2\,\logmsos^2 
              \right)
          \bigg]
       \nonumber\\&&\mbox{}
    + \left({M_t^2\over s}\right)^{3}\,\bigg[
           - {236639\over 1944} 
           + \left(
               - {7318\over 81} 
               - 16\,\ln 2
              \right)\,\zeta_2 
           + {280\over 9}\,\zeta_3
       \nonumber\\&&\mbox{\hspace{.5cm}}
           + {640\over 3}\,\logmsos 
           - {889\over 81}\,\logmsos^2 
           + n_l\,\left(
                 {269\over 243} 
               + {148\over 27}\,\zeta_2
               - {638\over 81}\,\logmsos 
               + {10\over 3}\,\logmsos^2 
              \right) 
          \bigg]
       \nonumber\\&&\mbox{}
    + \left({M_t^2\over s}\right)^{4}\,\bigg[
             {28244\over 729} 
           + \left(
               - {45389\over 162} 
               - 32\,\ln 2
              \right)\,\zeta_2 
           + {8\over 3}\,\zeta_3
           + {432461\over 972}\,\logmsos 
           + {4727\over 324}\,\logmsos^2 
       \nonumber\\&&\mbox{\hspace{.5cm}}
           + n_l\,\left(
               - {7061\over 1296} 
               + {40\over 3}\,\zeta_2
               - {1477\over 108}\,\logmsos 
               + {19\over 3}\,\logmsos^2 
              \right) 
          \bigg]
       \nonumber\\&&\mbox{}
    + \left({M_t^2\over s}\right)^{5}\,\bigg[
             {1248859307\over 2160000} 
           + \left(
               - {2008559\over 4050} 
               - 80\,\ln 2
              \right)\,\zeta_2 
           - 17\,\zeta_3
       \nonumber\\&&\mbox{\hspace{.5cm}}
           + {550105787\over 486000}\,\logmsos 
           - {337981\over 8100}\,\logmsos^2 
       \nonumber\\&&\mbox{\hspace{.5cm}}
           + n_l\,\left(
               - {6496853\over 243000} 
               + {3856\over 135}\,\zeta_2
               - {114617\over 4050}\,\logmsos 
               + {50\over 3}\,\logmsos^2 
              \right) 
          \bigg]
       \nonumber\\&&\mbox{}
    + \left({M_t^2\over s}\right)^{6}\,\bigg[
             {74282645263\over 29160000} 
           + \left(
               - {21379\over 30} 
               - 224\,\ln 2
              \right)\,\zeta_2 
           - {1136\over 15}\,\zeta_3
       \nonumber\\&&\mbox{\hspace{.5cm}}
           + {1145970713\over 486000}\,\logmsos 
           - {225373\over 540}\,\logmsos^2 
       \nonumber\\&&\mbox{\hspace{.5cm}}
           + n_l\,\left(
               - {2680259\over 30375} 
               + {9824\over 135}\,\zeta_2
               - {13901\over 225}\,\logmsos 
               + {1292\over 27}\,\logmsos^2 
             \right) 
          \bigg] 
\bigg\}
+ \ldots\,\,,
\label{eqr2a}
\\
%%%%%%%%%%%%%%%%%%%%%%%%%%%%%%%%%
  R^{(2),a}_S &=& 3\,\bigg\{
    \left({M_t^2\over q^2}\right)^{3}\,\left(
           - 74 
           + 24\,\zeta_3
           + 36\,\logmsos 
          \right)
       \nonumber\\&&\mbox{}
    + \left({M_t^2\over q^2}\right)^{4}\,\left(
           - {70\over 9} 
           - {88\over 3}\,\zeta_2 
           - 24\,\zeta_3
           + {205\over 9}\,\logmsos 
           + {44\over 3}\,\logmsos^2 
          \right) 
       \nonumber\\&&\mbox{}
    + \left({M_t^2\over q^2}\right)^{5}\,\left(
           - {4111\over 72} 
           - {830\over 9}\,\zeta_2 
           - 40\,\zeta_3
           + {670\over 3}\,\logmsos 
           + {415\over 9}\,\logmsos^2 
          \right) 
       \nonumber\\&&\mbox{}
    + \left({M_t^2\over q^2}\right)^{6}\,\left(
             {6044237\over 54000} 
           - {7771\over 15}\,\zeta_2 
           - 84\,\zeta_3
           + {1177331\over 900}\,\logmsos 
           + {7771\over 30}\,\logmsos^2 
          \right) 
\bigg\}
+ \ldots\,\,,\nonumber\\
\label{eqr2as}
\end{eqnarray}
where $\mu^2=s$ is chosen.
Note, that the quartic corrections of $\Pi_S^{(2),a}$ have no
imaginary parts so that $R_S^{(2),a}$ actually starts at order
$(M_t^2/s)^3$.
An important check of our result is provided by the successful
comparison of the terms proportional to $n_l$ with the
expansion of the exact analytical expression
\cite{HoaTeu96}.
The quartic terms for the proper sum $R^{(2),a}(s)$ are also
available in the literature
\cite{CheKue94}
and complete agreement was found.

For completeness we list the results from the double-triangle
diagrams containing cuts from the $b$ quark only
\cite{KniKue89}:
\begin{eqnarray}
R^{(2),a}_{Sb}(s) &=&
3\Bigg\{
-\frac{15}{8}
+\zeta_2
+\frac{M_t^2}{s}
\left[
2
-10\zeta_2
-6\logmsos
+\logmsos^2
\right]
\nonumber\\&&\mbox{}
+\left(\frac{M_t^2}{s}\right)^2
\left[
-\frac{39}{4} 
- \zeta_2
+ 8\zeta_3
+ \left(\frac{15}{2} - 2\zeta_2\right)\logmsos
+ \frac{1}{2}\logmsos^2 
+ \frac{1}{3}\logmsos^3
\right]
\nonumber\\&&\mbox{}
+\left(\frac{M_t^2}{s}\right)^3
\left[
\frac{91}{9} 
- 4\zeta_2
+ \frac{8}{3}\logmsos
+ 2\logmsos^2 
\right]
\nonumber\\&&\mbox{}
+\left(\frac{M_t^2}{s}\right)^4
\left[
\frac{1907}{144} 
- 5\zeta_2
+ \frac{95}{12}\logmsos
+ \frac{5}{2}\logmsos^2 
\right]
\nonumber\\&&\mbox{}
+\left(\frac{M_t^2}{s}\right)^5
\left[
\frac{75803}{2700} 
- \frac{28}{3}\zeta_2
+ \frac{826}{45}\logmsos
+ \frac{14}{3}\logmsos^2
\right]
\nonumber\\&&\mbox{}
+\left(\frac{M_t^2}{s}\right)^6
\left[
\frac{31073}{450} 
- 21\zeta_2
+ \frac{917}{20}\logmsos
+ \frac{21}{2}\logmsos^2 
\right]
\Bigg\}.
\end{eqnarray}
This contribution has to be subtracted from $R^{(2),a}_S$.  Note that
the cut arising from two gluons is zero according to the
Landau-Yang-Theorem
\cite{LanYan}.
\begin{figure}[tb]
 \begin{center}
 \begin{tabular}{cc}
   \leavevmode
   \epsfxsize=6.5cm
   \epsffile[110 270 480 560]{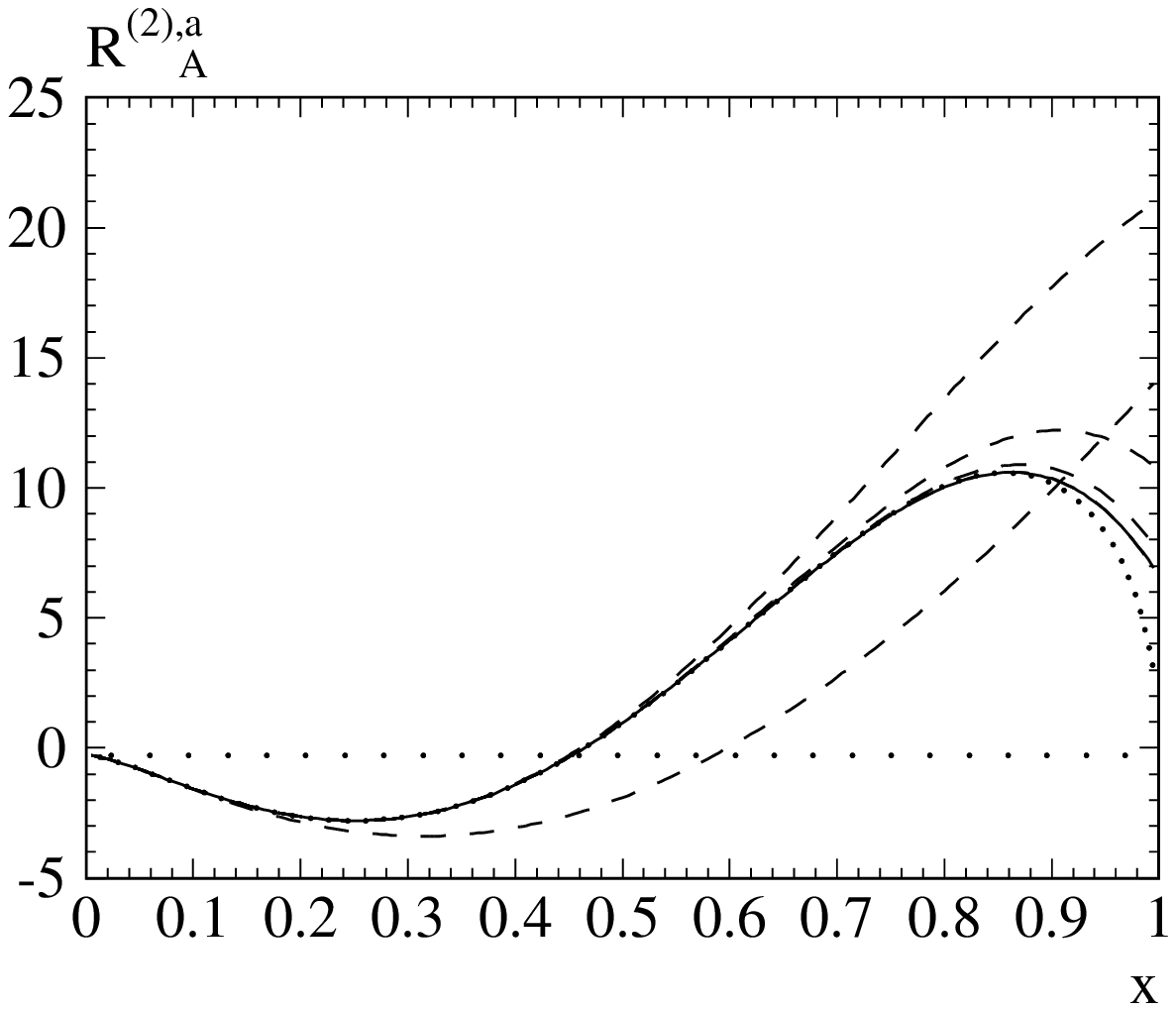}
   &
   \epsfxsize=6.5cm
   \epsffile[110 270 480 560]{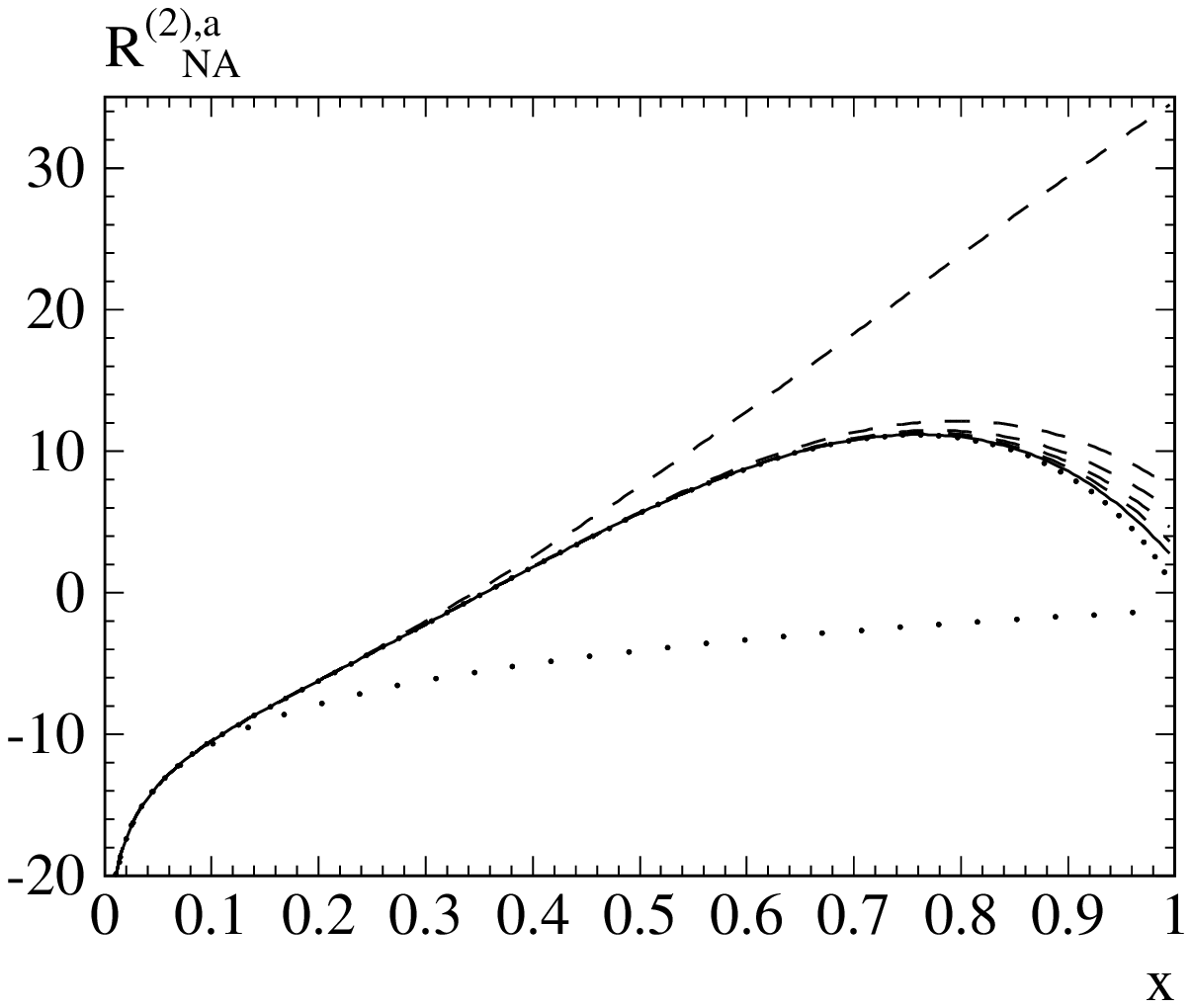} 
\\
   \epsfxsize=6.5cm
   \epsffile[110 270 480 560]{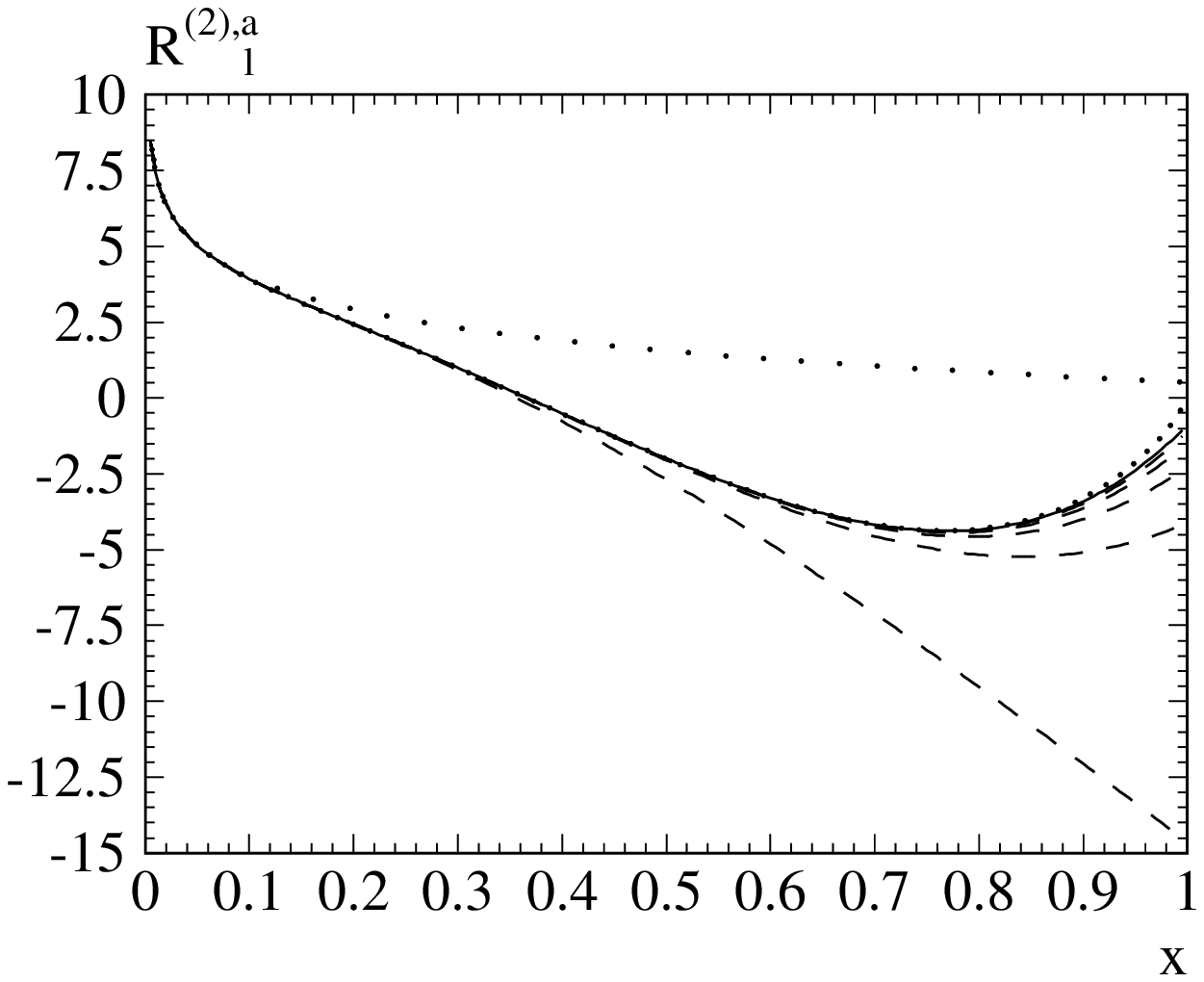}
   &
   \epsfxsize=6.5cm
   \epsffile[110 270 480 560]{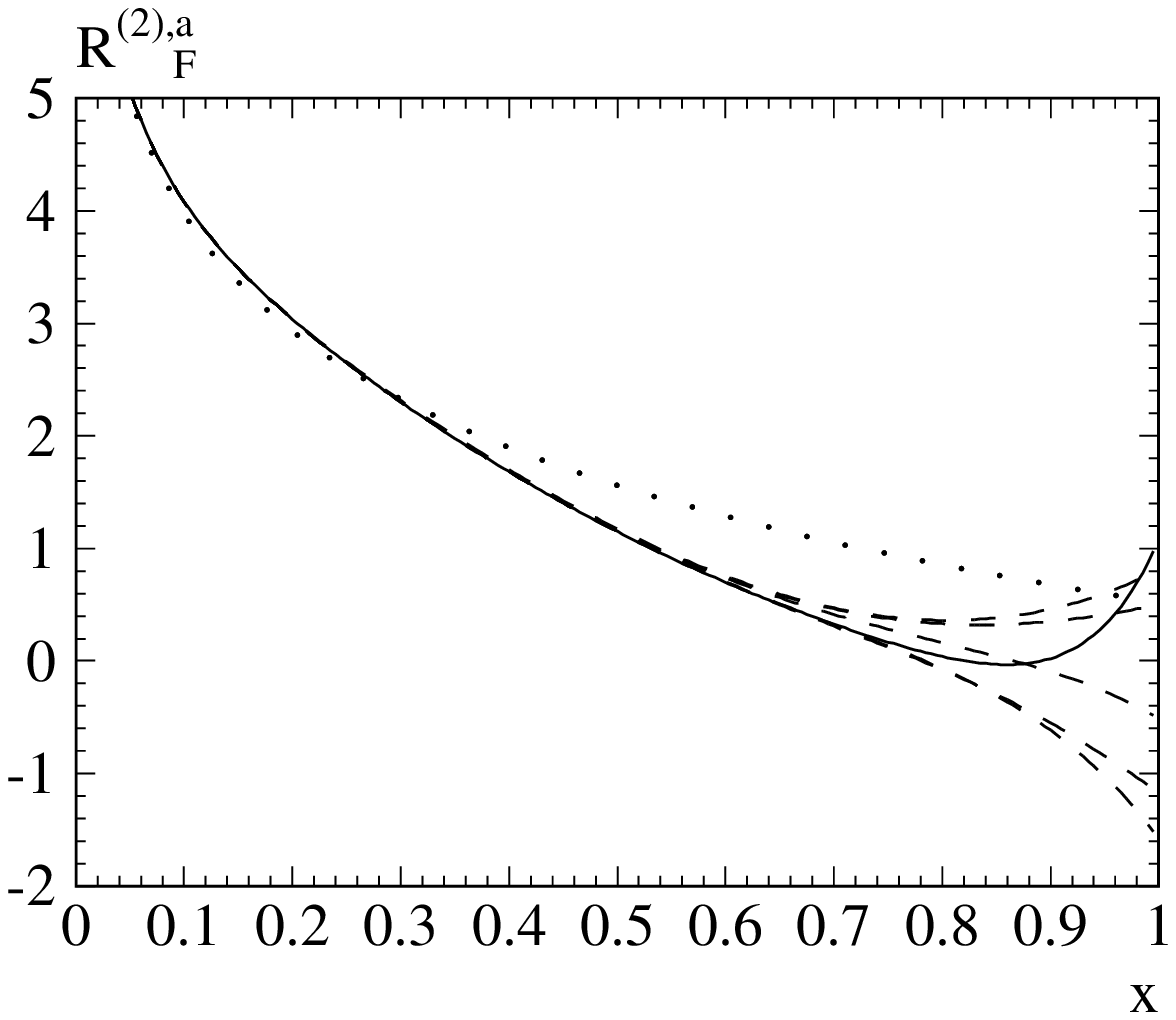}
\\
   \epsfxsize=6.5cm
   \epsffile[110 270 480 560]{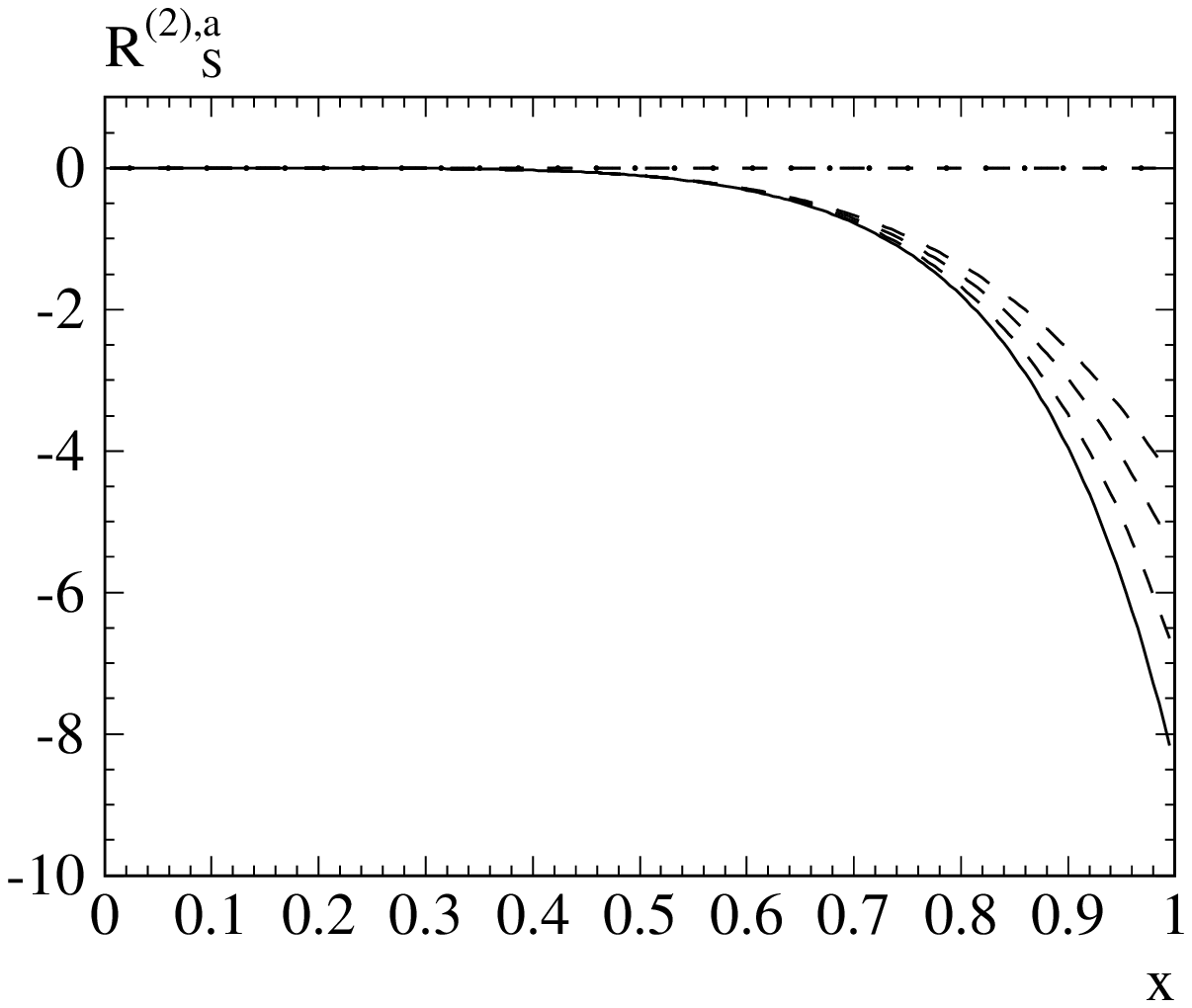}
   &
   \epsfxsize=6.5cm
   \epsffile[110 270 480 560]{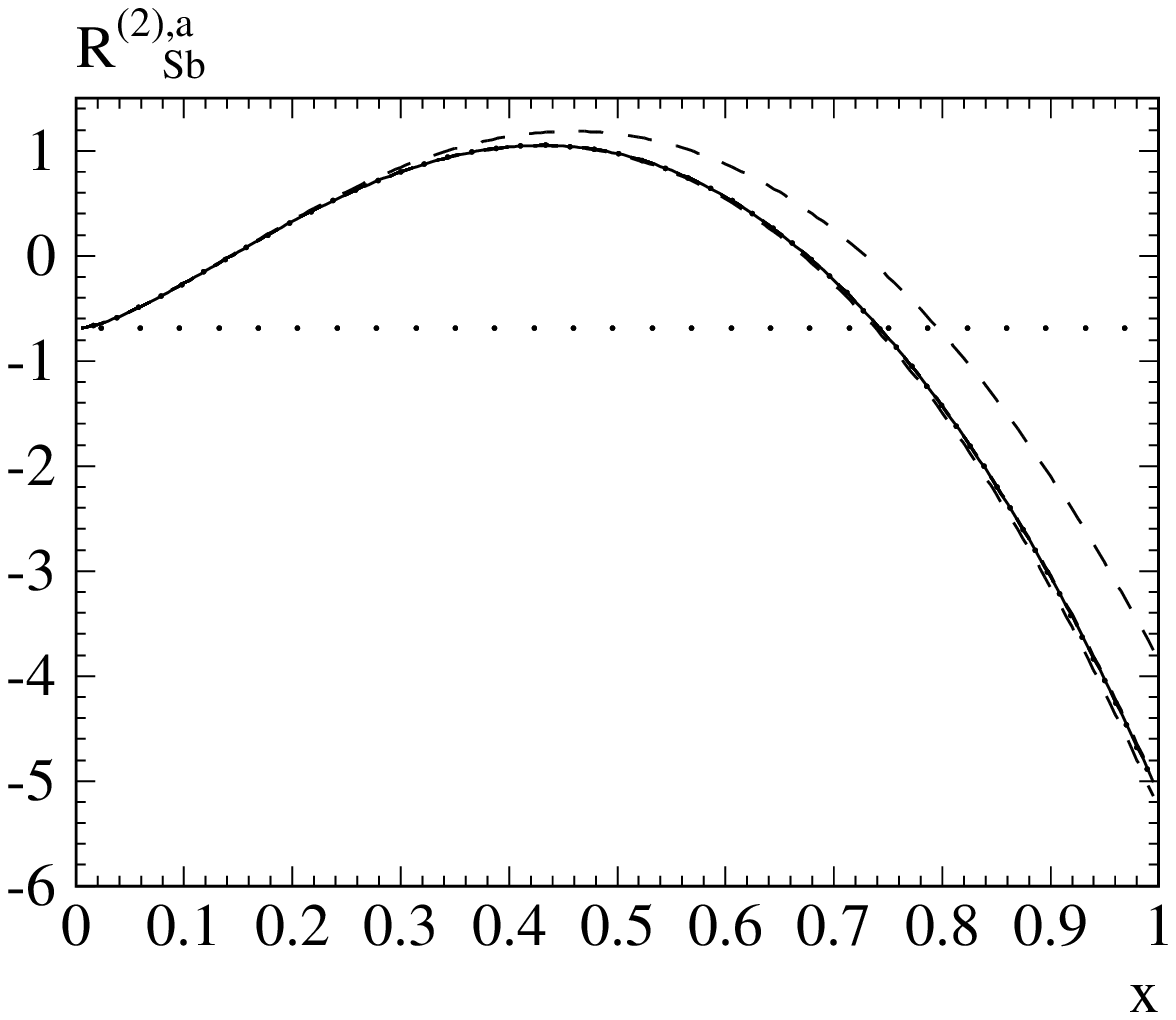}
 \end{tabular}
 \caption{\label{figra} 
   $R^{(2),a}_i$, $i={\it A, NA, l, F, S, Sb}$ as functions of $x =
   2M_t/\protect\sqrt{s}$ at $\mu^2 = M_t^2$. Successively higher order
   terms in $(M_t^2/s)^n$: Dotted: $n=0$; dashed: $n=1,\ldots,5$; solid:
   $n=6$. Narrow dots: exact result ($R_l^{(2),a}$, $R_{Sb}^{(2),a}$) or 
   semi-analytical results ($R_A^{(2),a}$, $R_{NA}^{(2),a}$).}
 \end{center}
\end{figure}
In Fig.~\ref{figra} the terms for the five different
contributions are plotted against $x=2M_t/\sqrt{s}$ including
successively higher orders in $M_t^2/s$. For $R_A^{(2),a}$ and
$R_{NA}^{(2),a}$ a comparison with a recently evaluated semi-analytical
result (narrow dots) \cite{CheKueSte97} is possible and agreement up to
$x\approx0.8$ is found.  The light fermion contribution, $R_l^{(2),a}$,
may be compared with exact results \cite{HoaTeu96} (narrow dots) and
also shows agreement up to $x\approx0.8$.  Concerning $R_F^{(2),a}$ and
$R_S^{(2),a}$ the situation is less satisfactory. It seems that there is
reasonable convergence up to $x\approx0.7$ which is also motivated by
the behaviour of the vector case where analytical results for $x>0.5$
are available (see \cite{CheHarKueSte97}).  However, for $x>0.7$ the
behaviour of the curve including all known power correction terms (solid
line) indicates that close to $x=1$ the convergence fails to work. The
reason presumably is connected to the four particle cut starting at
$x=0.5$.  Although $R_A^{(2),a}$ and $R_{NA}^{(2),a}$ also exhibit a
four particle cut it seems to be somehow less dominant for these
contributions.  In Fig.~\ref{figra} also the contribution
$R_{Sb}^{(2),a}$ is shown.  Here, already the curve including power
corrections up to order $(M_t^2/s)^{2}$ is practically indistinguishable
from the exact result. Note that only the difference of the two plots
in the bottom line of Fig.~\ref{figra} enters $R_t(s)$.

\begin{table}[t]
%%%{\footnotesize
\renewcommand{\arraystretch}{1.3}
\begin{center}
%%%%
\begin{tabular}{|l|l||r||r|r|r|r|}
 \hline 
 $\sqrt{s}$ (GeV) & $x$ &  $\alpha_s^{(6)}(s)$ & 
 $R_t^{(0)}$ &  $C_F R_t^{(1)}$ & 
 $R_t^{(2)}$ & $R_t(s)$ \\
  \hline\hline 
$      500   $ & $        0.70$ & $       0.095$ & $       1.419$ &
$       6.021$ & $      29.902$ & $       1.629$ \\
$     1000   $ & $        0.35$ & $       0.088$ & $       1.732$ &
$       2.842$ & $       6.016$ & $       1.816$ \\
$     1500   $ & $        0.23$ & $       0.085$ & $       1.771$ &
$       2.291$ & $       3.709$ & $       1.836$ \\
$     2000   $ & $        0.18$ & $       0.083$ & $       1.784$ &
$       2.091$ & $       3.044$ & $       1.841$ \\
\hline 
\end{tabular}
\end{center}
%%%}
\caption{\label{tabrt}Numerical values for the contributions
of ${\cal O}(\alpha_s^i)$, $R_t^{(i)}$, to the normalized cross
section $R_t$. The values for $\alpha_s^{(6)}(s)$ are based on
$\alpha_s^{(5)}(M_Z^2)=0.118$. The scale $\mu^2=s$ has been adopted.
Also the values of $x=2M_t/\protect\sqrt{s}$ are shown. 
}
\end{table}

\begin{figure}[t]
  \begin{center}
    \leavevmode
    \epsfxsize=8.cm
    \epsffile[110 185 465 630]{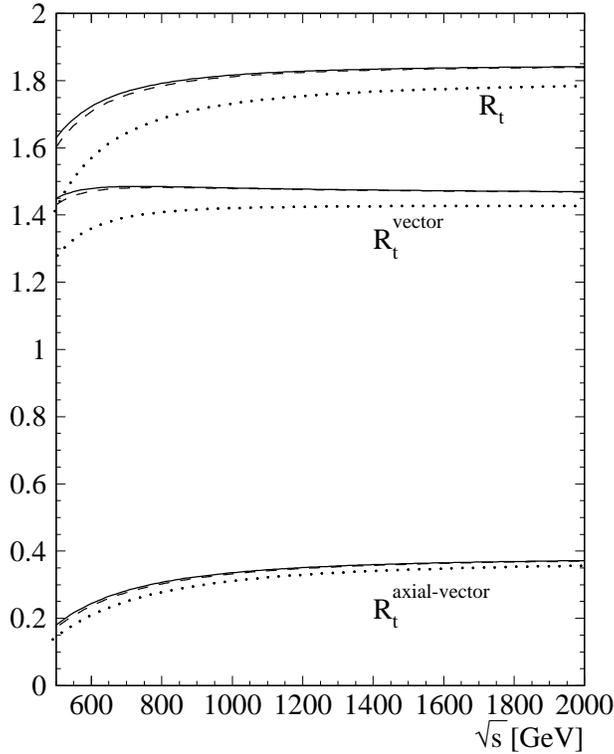}
    \hfill
    \parbox{14.cm}{\small
    \caption[]{\label{figrt}\sloppy\rm
      Normalized cross section $R_t$ as a function of the center of mass
      energy $\sqrt{s}$, together with the pure vector and axial-vector
      contributions: $R_t = R_t^{\rm vector}+R_t^{\rm axial-vector}$.
      Dotted: Born approximation; dashed: ${\cal
        O}(\alpha_s)$, solid: ${\cal O}(\alpha_s^2)$. The scale
      $\mu^2=s$ has been adopted.}}
  \end{center}
\end{figure}
\begin{figure}
  \begin{center}
    \leavevmode
    \epsfxsize=10.cm
    \epsffile[110 265 465 560]{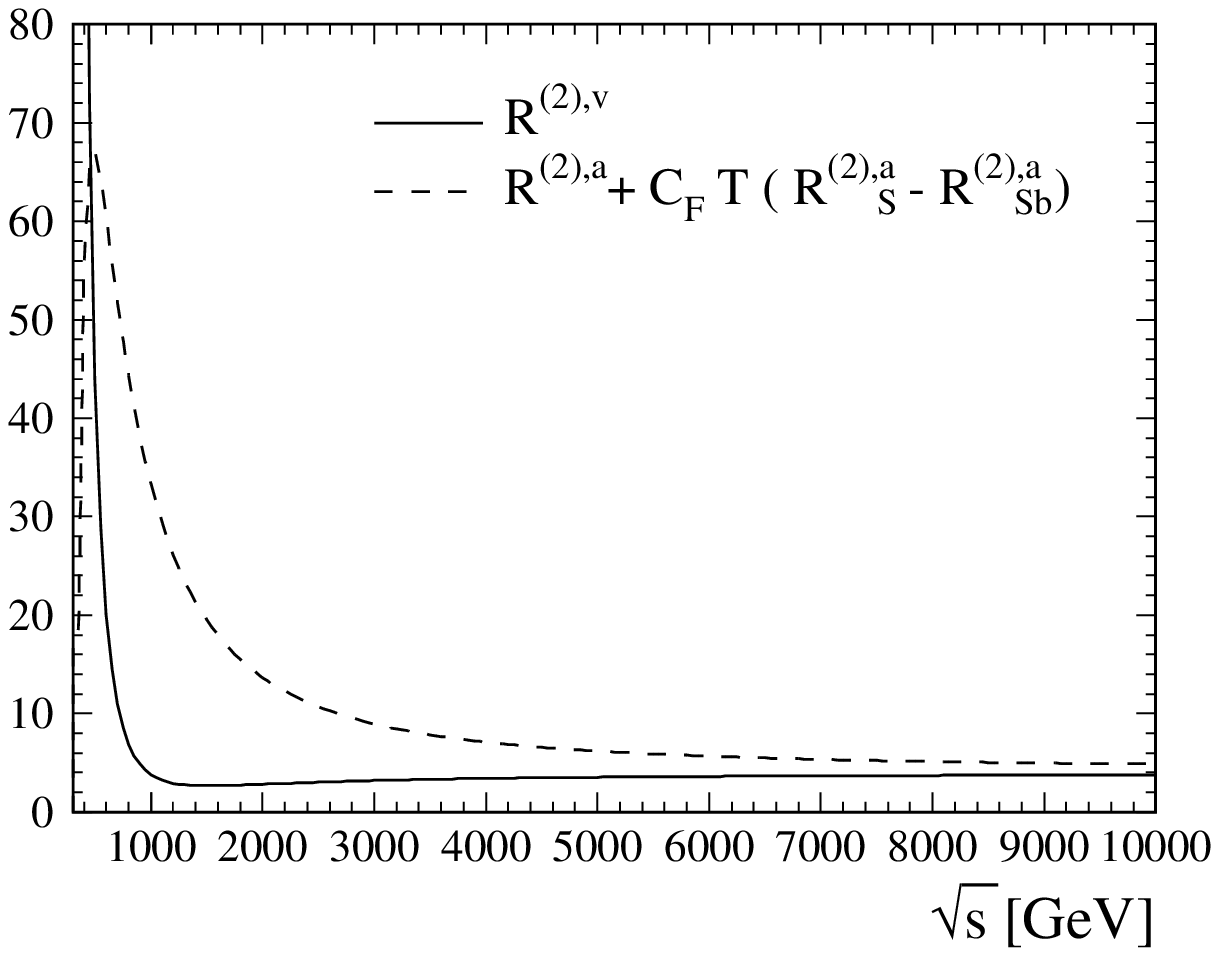}
    \hfill
    \parbox{14.cm}{\small
      \caption[]{\label{figr2av}\sloppy\rm
        Two-loop vector and axial-vector contributions $R^{(2),v}$,
        $R^{(2),a} + C_F T R^{(2),a}_S$. The scale $\mu^2=s$ has been 
        adopted and $M_t=175$~GeV has been chosen.}}
  \end{center}
\end{figure}
Recalling that the first seven terms for $R^v(s)$ approximate the
exact result up to $x\approx0.7$
\cite{CheHarKueSte97}
we are now prepared to present 
predictions for $R_t$ valid up to ${\cal O}(\alpha_s^2)$ 
for $\sqrt{s}\gsim500$~GeV which corresponds to $x\lsim0.7$.
Therefore we insert the isospin and charge 
quantum numbers into Eq.~(\ref{eqtt}) and choose $n_l=5$, 
$s_\Theta^2=0.2315$,
$\alpha_s^{(5)}(M_Z^2)=0.118$, $M_Z=91.187$~GeV and $M_t=175$~GeV. 
Then the expansion of $R_t(s)$ 
looks as follows:
\begin{eqnarray}
R_t(s) &=& R_t^{(0)}(s) 
+\frac{\alpha_s^{(6)}(s)}{\pi}\,C_F\,R_t^{(1)}(s) 
+\left(\frac{\alpha_s^{(6)}(s)}{\pi}\right)^2\,R_t^{(2)}(s)
\end{eqnarray}
In Tab.~\ref{tabrt} the coefficients $R_t^{(i)}$ are listed for
different values of the center of mass energy $\sqrt{s}$.
One observes that for $\sqrt{s}=500$~GeV, which is a proposed option for 
the NLC, the ${\cal O}(\alpha_s^2)$ QCD corrections amount to $\approx2\%$.
For higher values of the center of mass energy these terms get less
important.
In Fig.~\ref{figrt} the normalized cross section $R_t$ is plotted
against $\sqrt{s}$. 
The contributions from the vector and axial-vector part are also
displayed separately. 
$R_t(s)$ is clearly dominated by the vector
contribution which is mainly due to the fact that in Eq.~(5) the
couplings to $R^v$ are larger by roughly a factor of four as compared to
$R^a$. Another reason is that the Born cross section $R^{(0),v}$ is
always larger than $R^{(0),a}$. This is not true for the ${\cal
  O}(\alpha_s)$ and ${\cal O}(\alpha_s^2)$ terms. Here the axial-vector
contribution exceeds the vector part for sufficiently large values of
$\sqrt s$ and approaches it from above as $\sqrt s$ goes to infinity,
where both $R^v$ and $R^a$ are identical.  In Fig.~\ref{figr2av} this is
demonstrated at order $\alpha_s^2$. For this reason at energies above
roughly $\sqrt s = 600$~GeV the tree-loop vector and axial-vector
contributions to $R_t$ are comparable.  
At lower values of the energy
the vector part is still larger than the axial-vector part as a
consequence of the more singular threshold behaviour of $R^v$
(Fig.~\ref{figr2av}).  For the sake of completeness we note that the
contribution from the singlet diagram which is absent in the vector case
is smaller by at least a factor $100$ as compared to the non-singlet
case.

To conclude, the large momentum procedure has been applied to the
axial-vector polarization function and terms up to order $(M_t^2/q^2)^6$
have been determined. The imaginary part in combination with the result
recently obtained for the vector case was used to predict the production
of top quarks at future $e^+ e^-$ colliders up to ${\cal
  O}(\alpha_s^2)$.

\vspace{2mm}
\centerline{\bf Acknowledgments}
\smallskip\noindent
We would like to thank K.G. Chetyrkin and J.H. K\"uhn
for valuable comments and carefully reading the manuscript.
M.S. thanks B.A. Kniehl for discussions in connection with $R_{Sb}$.

\vspace{-1.5mm}

%%%%%%%%%%%%%%%%%%%%%%%%%%%%%%%%%%%%%%%%%%%%%%%%%%%%%%%%%%%%
%%%%%%%%%%%%%%%%%%%%%%%%%%%%%%%%%%%%%%%%%%%%%%%%%%%%%%%%%%%%

\end{document}